\documentclass[12pt,preprint]{aastex}
\newcommand{\cm}{$\,{\rm cm}$}
\newcommand{\kpc}{$\,{\rm kpc}$}

\newcommand{\persqcm}{$\,{\rm cm^{-2}}$}

\newcommand{\kms}{\ensuremath{\,{\rm km\,s}^{-1}}}

\newcommand{\hi}{{\rm H\,}{{\sc i}}}
\newcommand{\hii}{{\rm H\,}{{\sc ii}}}

\newcommand{\cor}{\ensuremath{^{13}{\rm CO}}}

\newcommand{\expo}[1]{\ensuremath{10^{#1}}}

\newcommand{\hisa}{{\rm H\,}{{\sc i}} SA}
\newcommand{\xmm}{{\it XMM-Newton}}

\newcommand{\fdeg}{\hbox{$.\mkern-4mu^\circ$}}   
\newcommand{\oqkev}{$1\over4$~keV}
\newcommand{\tqkev}{$3\over4$~keV}
\newcommand{\kev}{{\rm keV}}
\newcommand{\irdci}{G28.37+0.07}
\newcommand{\irdcii}{G36.67$-$0.11}
\shorttitle{X-ray Shadowing of IRDCs}
\shortauthors{Anderson, Snowden, \& Bania}
\begin{document}

\title{X-ray Shadowing Experiments Toward Infrared Dark Clouds}

\author{L. D. Anderson\altaffilmark{1,2}, S. L. Snowden\altaffilmark{3} \& T. M. Bania\altaffilmark{1}}

\altaffiltext{1}{Institute for Astrophysical Research, Department of Astronomy, 725 Commonwealth Ave., Boston University, Boston MA 02215, USA.}
\altaffiltext{2}{Current address: Laboratoire d'Astrophysique de Marseille, 38 rue F. Joliot-Curie, 13388 Marseille Cedex 13, France}
\altaffiltext{3}{NASA Goddard Space Flight Center, Code 662, Greenbelt, MD 20771, USA.}

\begin{abstract}
  We searched for X-ray shadowing toward two infrared dark clouds
  (IRDCs) using the MOS detectors on {\it XMM-Newton\/} to learn about
  the Galactic distribution of X-ray emitting plasma.  IRDCs make
  ideal X-ray shadowing targets of \tqkev\ photons due to their high
  column densities, relatively large angular sizes, and known
  kinematic distances.  Here we focus on two clouds near $30\arcdeg$
  Galactic longitude at distances of 2 and 5\kpc\ from the Sun.  We
  derive the foreground and background column densities of molecular
  and atomic gas in the direction of the clouds.  We find that the
  \tqkev\ emission must be distributed throughout the Galactic disk.
  It is therefore linked to the structure of the cooler material of
  the ISM, and to the birth of stars.
\end{abstract}


\section{Introduction}
The diffuse X-ray background of the Milky Way has been studied for
over 30 years and yet relatively little is known about the
distribution of its emission in the Galactic plane.  Most diffuse
Galactic emission arises from $\sim1-3\times10^6$\,K plasma
\citep{mccammon83}, but the origin and distribution of this plasma are
still debated.  Two factors are responsible for the difficulty in
determining the distribution of this plasma: confusion and absorption.
Confusion arises because one cannot determine the origin of an
individual X-ray.  Even though most X-rays contributing to the diffuse
background of the Milky Way are thermal in origin and the spectra of
this plasma show strong line emission, the available X-ray detector
energy resolution is insufficient to provide velocity (and therefore
distance) information.  Thus it is very difficult to determine the
true distribution of X-ray emitting plasma along a given line of
sight.


Shadowing experiments are the only way to determine the distribution
of hot plasma in the plane of the Milky Way.  To date, they have been
used primarily in the \oqkev\ band to study the distribution of
0.1~keV plasma within the Local Hot Bubble and in the lower halo
\citep[e.g.,][]{snowden00}.  A few shadowing experiments at higher
energy (\tqkev\ and 1.5~keV) using {\it ROSAT} data
\citep[e.g.,][]{park97, almy00} have shown evidence for the existence
of an extensive distribution of hot plasma well within the solar
circle.  This emission has been linked to the Milky Way X-ray bulge
which has a scale height of $\sim1.8$~kpc and a radial extent of
$\sim5$~kpc \citep{snowden97}, but there are likely additional
components distributed throughout the Galactic disk.

Most studies of the X-ray background have focused on the Local Hot
Bubble (LHB).  The LHB is a region of about 100\,pc in radius,
surrounding the Sun filled with a hot, rarefied $10^6$\,K plasma
\citep[e.g.,][]{snowden90}.  Beginning with \citet{snowden93}, many 
groups have effectively mapped the distribution of this gas
\citep[e.g.][]{smith05, galeazzi07, henley07, smith07, henley08}.
These studies have shown the utility of X-ray shadowing in mapping a
large region of the sky at higher Galactic latitudes.  Here we utilize 
similar techniques but extend the analysis to larger distances in the
Galactic plane using dense molecular clouds.

Infrared Dark Clouds (IRDCs) are very dense molecular clouds that appear
dark at infrared wavelengths. They are seen in absorption against the
mid-IR background.  IRDCs were first identified
as a significant population in the {\it Midcourse Space Experiment}
({\it MSX}) mid-infrared data \citep{egan98}, and are a ubiquitous
feature of the higher resolution Galactic Legacy Infrared Survey
Extraordinaire
\citep[GLIMPSE:][]{benjamin03}. IRDCs have very high column densities of
$\expo{22} \cm^{-2}$ \citep[][hereafter S06]{simon06}, and possibly as
high as $\expo{23} - \expo{25} \cm^{-2}$ \citep{egan98, carey98}.
They exist at a wide range of distances and are most easily seen in
the plane of the Galaxy (S06).

S06 found distances to 313 IRDCs using the Galactic Ring Survey of
\cor\ emission \citep[GRS; ][]{jackson06} to establish a morphological
match between the \cor\ gas and the mid-infrared extinction.  Using
the rotation curve of \citet{clemens85}, S06 converted the velocity of
the associated \cor\ gas into a distance to the IRDC.  In the inner
Galaxy (the focus of the present study), every velocity has two
possible distance solutions.  IRDCs, however, are seen in absorption
against the Galactic plane and therefore the near distance can be
assumed.

IRDCs make ideal X-ray shadowing candidates.
Their high column densities make them effective absorbers of
background X-ray photons; X-ray emission from the direction of the
IRDC must originate in the foreground.  Thus, using IRDC absorption,
one can separate foreground and background X-ray components.  The
distribution of IRDCs also spans a wide range of distances \citep[see
S06,][]{jackson08}.  With enough X-ray absorption measurements of
IRDCs along different lines of sight, an accurate map of the hot ISM
may be created.

\section{IRDC Sample}
The present study includes two clouds from the catalog of S06 that
have high column densities, large angular sizes, and are closely
grouped on the sky. Our goal is to select clouds that would cause
complete absorption of X-ray photons originating behind the cloud.
The models given in \citet{snowden94} predict that a total column
density of $1 \times \expo{22} \cm^{-2}$ will absorb $\sim\,98\%$ of
incident photons at \tqkev.
Column densities of $10^{22}$ are frequently found for IRDCs (see S06).
To ensure sufficient counts in both the on-- and off--cloud
directions, the target IRDCs should cover approximately half the \xmm\
detector.  Our ideal situation is one in which there are multiple
large clouds closely grouped on the sky but at various distances.  The 
close grouping allows one to disentangle the emission components along 
a given line of sight.

The properties of the two IRDCs observed by \xmm\ are summarized in
Table~\ref{tab:prop}.  Listed are the IRDC name, the Galactic
longitude and latitude, the angular size, and the molecular column
density.  All parameters in Table~\ref{tab:prop} are reproduced from
S06.  The two target clouds have high column densities, large
angular sizes, and are near $l = 30 \arcdeg$.  

\begin{figure}
\includegraphics[width=6.5in]{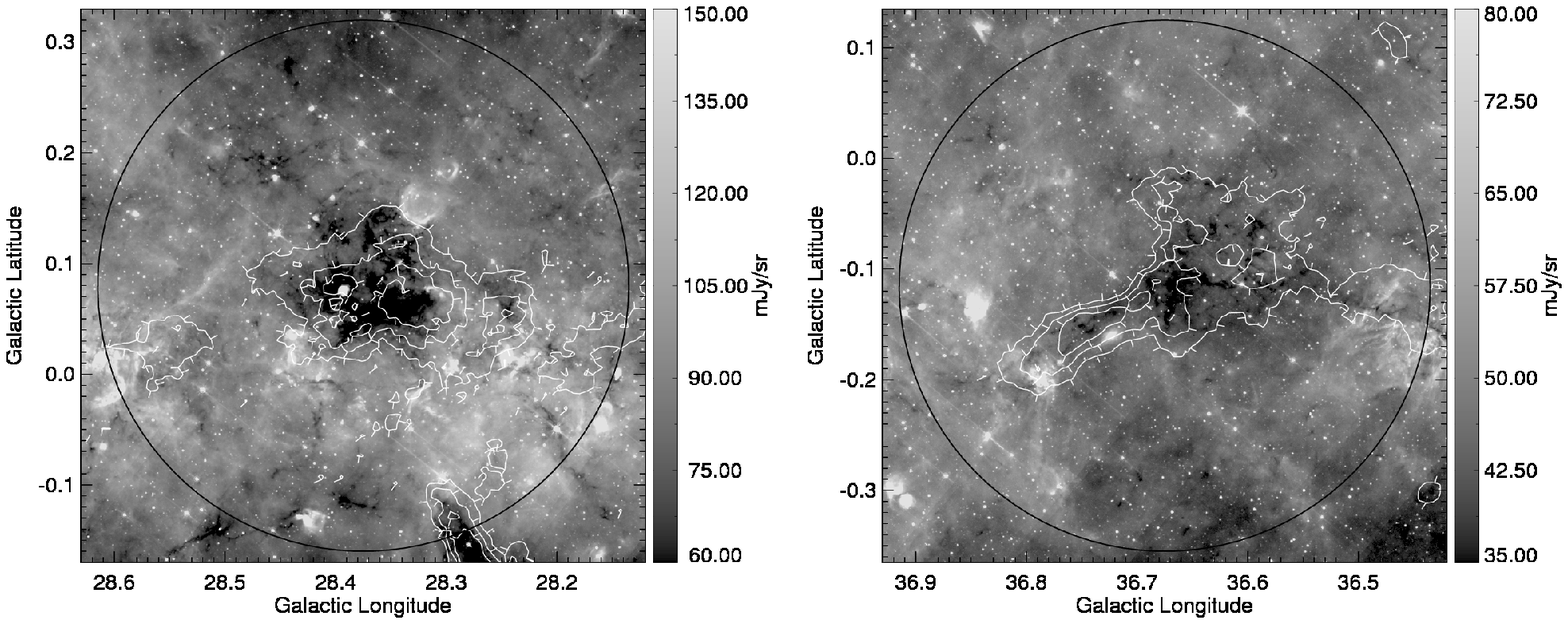}
\caption{IRDCs shown in absorption against GLIMPSE 8\micron\ emission. The 
left panel shows \irdci\ and the right panel shows \irdcii.  The
contours are \cor\ integrated intensity from the GRS.  Tick-marks on
the contours point downhill, toward decreasing \cor\ integrated
intensity.  The large black circles represent the \xmm\ field of view
($\sim 30 \arcmin$).}

\label{fig:clouds}
\end{figure}

The mid-infrared emission of the two IRDC fields is shown in
Figure~\ref{fig:clouds}.  The background image in this figure is from
the GLIMPSE 8\micron\ survey \citep{benjamin03} and the contours are
\cor\ integrated intensity from the GRS.  We have created the
integrated intensity map using the LSR velocity and line width from
the S06 catalog.  Tick-marks on the contours indicate the direction of
decreasing \cor\ integrated intensity.  The large circle in this
figure is approximately the \xmm\ field of view ($\sim30\arcmin$).
Immediately evident in this figure is the excellent morphological
match between the GRS \cor\ contours and the strong mid-infrared
extinction.  The molecular clouds associated with the IRDCs take up a
significant fraction of the \xmm\ field of view.  IRDC \irdci\ is
located at the center of a large, dense molecular cloud which is
roughly ellipsoidal in projection.
The two clouds are at very different
distances, \irdci\ lies $\sim 5$ kpc from the Sun and \irdcii\ lies
$\sim 2$ kpc from the Sun (S06).

\section{Observations and Data Reduction}
Our clouds were observed by \xmm\ on 2006 April 11 for 25
ks each (IRDC \irdci\ with \xmm\ ObsID 0302970301 and IRDC \irdcii\ 
with ObsID 0302970201).

One of us (SLS) has developed a powerful suite of {\it PERL} and {\it
  FORTRAN} scripts to analyze \xmm\ extended source data called
XMM-ESAS \citep{snowden08}.  This software is based on the background
modeling described in \citet{snowden04} and \citet{kuntz08}.  The
software currently only operates on the MOS detector data and
subsequent to this analysis has been incorporated into the \xmm\
mission Standard Analysis Software (SAS) package.

To remove intervals of high soft proton contamination, we must first
temporally filter and clean the events file.  The signature of soft
proton contamination is a fluctuating light curve.  The XMM-ESAS
software creates a count rate histogram and fits a Gaussian to the
peak of the histogram.  In observations with minimal contamination,
this histogram would appear Gaussian; a non-Gaussian shape to the
distribution is also indicative of soft-proton contamination.  The
software defines count-rate intervals within $1.5 \sigma$ of the
median count rate as acceptable, and filters the events file to remove
intervals that do not satisfy this criterion.  While effective at removing intermittent
contamination, this process does not remove consistent low-level
contamination that is also likely present.  The light-curve for
\irdci\ is quite steady and almost all the data are useable.  The
light-curve of \irdcii, however, indicates a high level of soft proton
contamination and only $\sim 20 \%$ of the observation time is usable.

\section{Estimation of Column Density \label{sec:column}}
An estimate of the total column density to each IRDC can drastically
improve the reliability and accuracy of the X-ray spectral fitting.  Above
0.5 \kev, the most significant source of X-ray absorption is metals.
With normal abundances, the total X-ray opacity should be proportional to
$N({\rm H}) + 2N({\rm H_2})$.  Thus, we estimate the molecular column
density using the \cor\ emission from the GRS, and estimate the atomic
column density using the 21 cm \hi\ VLA Galactic Plane Survey
\citep[VGPS: ][]{stil06}.

Our spectral fitting procedure treats the on-- and off--cloud spectra
separately (as do all X-ray shadowing experiments).  This is the only
way one can separate components of the X-ray background that originate
in front of the cloud from those that originate behind it.  One must
therefore compute on-- and off--cloud column densities separately.  We
do this by first defining the on-- and off--cloud regions.

One of us (LDA) has developed an astronomical software program in
IDL.\footnote[1]{Download at http://www.bu.edu/iar/kang/.}  This
software allows one to define regions of interest in an image using a
threshold selection tool.  The user can vary the threshold upwards or
downwards to define smaller or larger regions.  The output of this
thresholding selection, the {\it x} and {\it y} positions of the data
points on the threshold boundary in sky coordinates, can then be
converted to detector coordinates within the software, as is
appropriate for analysis in the \xmm\ Science Analysis Software
(SAS\footnote[2]{http://xmm.esac.esa.int/sas/}).  This conversion
employs the SAS program {\it edet2sky} repeatedly for each coordinate
pair to convert from a region in sky coordinates to a region in
detector coordinates.

Using our IDL software, we define the on--cloud region as the sky
positions possessing both high mid-infrared extinction and high \cor\
column density.  Our goal is to create a region large enough to have
sufficient counts in the on--cloud region for good statistics.
Ideally, the on-- and off--cloud regions would each cover half the
\xmm\ field of view.  The off--cloud region encompasses all locations
in the \xmm\ field of view not including the on--cloud region. The
on--cloud region roughly corresponds to the largest contour in
Figure~\ref{fig:clouds}.

\subsection{Molecular Column Density}
\begin{figure}
\includegraphics[width=6.5in]{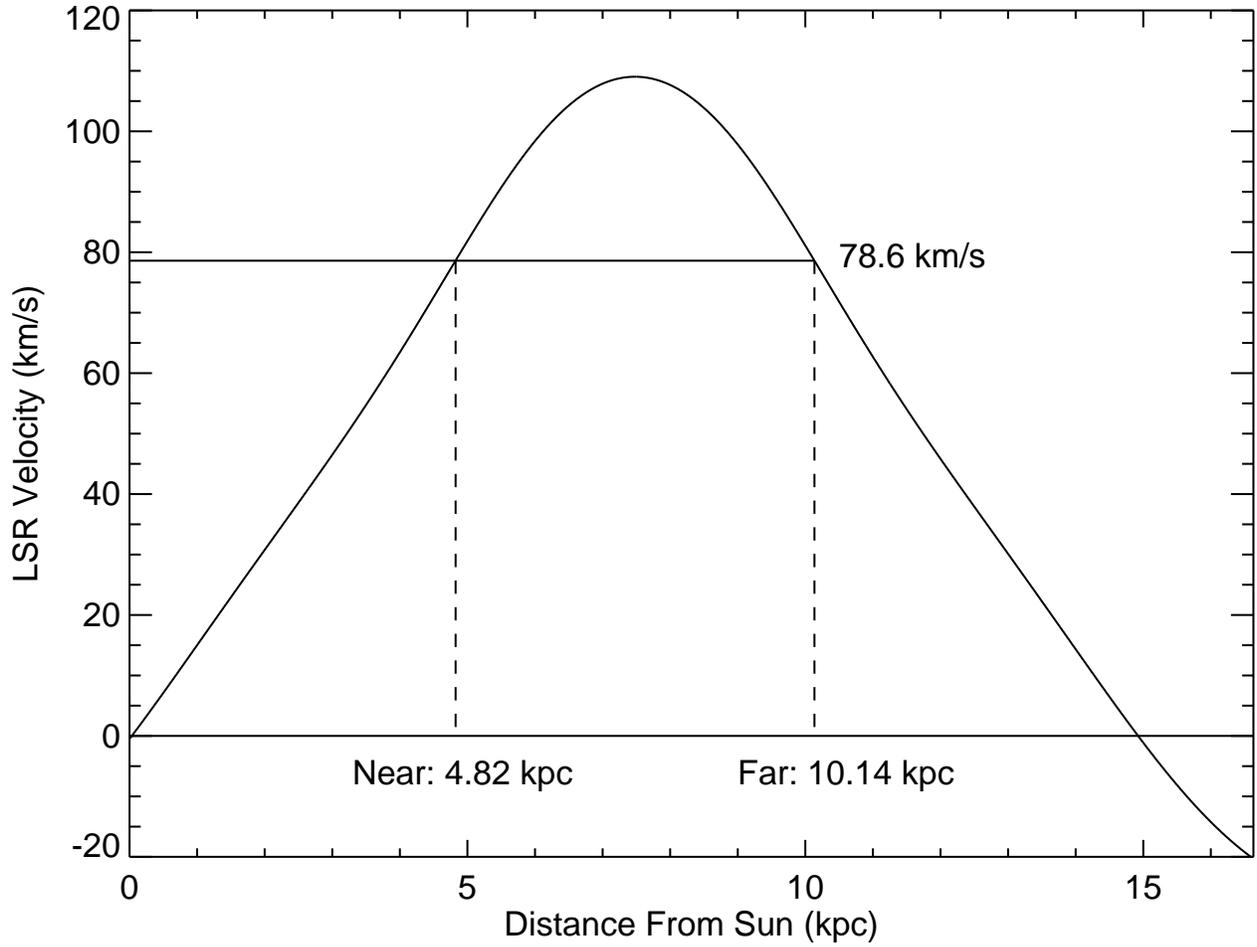}

\caption{Velocity versus distance plot for \irdci\ showing the
 kinematic distance ambiguity.  This cloud lies at the near distance
 of 4.82 kpc.}

\label{fig:kda}
\end{figure}

We must separate the total molecular column density into a component
foreground to the IRDC and a component beyond the IRDC.  We therefore
must estimate the distances to the molecular gas clumps that
contribute to the column density in the \xmm\ field of view.  In the
first Galactic quadrant (where the target IRDCs lie), each radial
velocity has two possible distance solutions, a ``near'' and a ``far''
distance.  This problem is known as the kinematic distance ambiguity
(KDA).  Gas along the same line of sight as an IRDC could lie either
in front of or behind the cloud.  This problem is illustrated in
Figure~\ref{fig:kda} for IRDC \irdci.  Figure~\ref{fig:kda} shows the
loci of distances from the Sun corresponding to a particular velocity,
assuming the Galactic rotation curve of \citet{clemens85}.  The
two-fold distance degeneracy is clearly shown.  Because IRDCs are seen
in absorption, we assume they lie at the near distance.  The cloud
velocity of 78.6 \kms\ places \irdci\ at a distance of 4.8 \kpc\ from
the Sun.  Any gas with a velocity greater than 78.6 \kms\ must lie
beyond the cloud.  Molecular gas at velocities less than 78.6 \kms,
however, could lie at either the near or the far distance.

\hi\ self-absorption (\hisa) can remove the distance ambiguity of the
molecular gas clumps in the \xmm\ field of view.  This technique
relies on cold foreground \hi\ absorbing the emission of warm
background \hi\ at the same velocity.  Warm \hi\ is ubiquitous in our
Galaxy and emits at all allowed velocities.  \citet{litzt81}
hypothesized that molecular clouds must contain \hi, a result which
has been confirmed in numerous subsequent observations
\citep[e.g.,][]{kuchar93, williams96}.  This small population of cold
neutral \hi\ atoms inside molecular clouds is maintained by
interactions with cosmic rays.  The \hi\ inside molecular clouds at
the near distance will produce an absorption signal because there is
ample warm \hi\ emitting at the same velocity at the far distance.
Any cloud at the far distance should not show \hisa\ because there is
no background \hi\ emitting at the same velocity.  \hisa\ is therefore
a general technique for finding distances to molecular clouds.  This
technique was shown by \citet{jackson02} to be effective at resolving
the KDA for a dense molecular cloud, by \citet{busfield06} for
resolving the KDA for young stellar objects, and by
\citet{anderson09a} for resolving the KDA for the molecular gas
associated with \hii\ regions, and by \citet{julia09} for resolving
the KDA for GRS molecular clouds.

We utilize the procedure outlined in \citet{simon01} to convert from
GRS \cor\ emission line parameters into an estimate of column density.
This procedure assumes a standard CO excitation temperature of 10\,K.  
The conversion equation is:
\begin{equation} \label{eqn:n_co} N({\rm H_2}) = 5.2 \times 10^{20} \, T \Delta V\,[{\rm cm^{-2}}]\,, \end{equation} 
where $T$ is the main beam line intensity in degrees Kelvin and $\Delta V$
is the line width in \kms.  The column density found using this
procedure is accurate to within a factor of $\sim 2$ due to
uncertainties in the conversion from \cor\ to ${\rm H_2}$.

To find the molecular column densities for both on-- and off--cloud
directions, we calculate the column density of all significant
\cor\ molecular clumps in the \xmm\ field of view by first fitting
Gaussians to the average on-- and off-- cloud GRS spectra.  We
transform the line parameters derived from these Gaussians into column
densities using Equation~\ref{eqn:n_co}.  The application of
Equation~\ref{eqn:n_co} computes the total column density along each line
of sight, but gives no information about whether the components lie at
the near or the far distance.  By examining the average \cor\ and
\hi\ spectrum of each molecular clump in the \xmm\ field of view for
\hisa, we determine whether the near or the far distance is
appropriate for each molecular clump.  We assign the column densities
of the molecular clumps found to lie in front of the target cloud to
the foreground component, and the column densities of the molecular
clumps found to lie behind to the target cloud to the background
component.

\subsection{Atomic Column Density}
We use a similar procedure to calculate the \hi\ column density using
the VGPS.  The conversion to \hi\ column density employs the equation:
\begin{equation} \label{eqn:n_hi} N({\rm H\,I}) = 1.82 \times 10 ^{18} \  
T_{\rm s} \int \tau(v) \, {\rm d}v \,[{\rm cm^{-2}}]\,, \end{equation} where $T_{\rm s}$
is the \hi\ spin temperature in degrees Kelvin, and $v$ is given in \kms.  
For \hi, the spin temperature is much less well constrained
compared to the excitation temperature of \cor.  A good average value
between the hot and cool \hi\ components is 150\,K \citep[see][and references therein]{dickey90}.  Equation~\ref{eqn:n_hi}, 
requires an estimate of the optical depth for the
\hi\ line using the equation:
\begin{equation} \label{eqn:tau_hi} \tau_{\rm HI} = -{\rm ln}
\left(1 - \frac{T_{\rm L}}{T_{\rm s} - T_{\rm
    BG}}\right)\,,\end{equation} where $\tau_{\rm HI}$ is the optical
depth of \hi, $T_{\rm L}$ is the \hi\ line intensity, $T_{\rm s}$ is
the \hi\ spin temperature, and $T_{\rm BG}$ is the intensity from
background sources.  The largest source of background emission is the
cosmic microwave background; because it produces a negligible effect
at 21\,cm, we ignore its contribution here.  Assuming a spin
temperature of 40\,K lowers our calculated column densities by a
factor of $\sim 2$.

For \hi\, there is less information about whether the gas originates
in front of or behind the cloud compared to CO.  As we did for \cor, we compute
average on-- and off--cloud spectra.  A typical \hi\ spectrum does not
show distinct, clean lines.  We therefore compute the area under the
curve directly, instead of fitting Gaussians, and convert this to
\hi\ column density using Equations~\ref{eqn:n_hi} and
\ref{eqn:tau_hi}.  We assume that the \hi\ is uniformly distributed
along the line of sight to a distance of 15\,kpc from the Sun and
split the column density distribution into foreground and background
components, weighted by the distance to the cloud.  For example, we
assume \irdci\ (at a distance of 5\,kpc) has 33\% of the total
\hi\ along the line of sight in the foreground component, and 67\% in
the background component.  Our values for the total \hi\ column
density are in very good agreement with those in the the
Leiden/Argentine/Bonn \hi\ survey \citep{kalberla05} and the
\hi\ survey of \citet{dickey90} as found using the HEASARC online
tool.\footnote[3]{http://heasarc.gsfc.nasa.gov/cgi-bin/Tools/w3nh/w3nh.pl}

\begin{figure}
\includegraphics[width=6.5in]{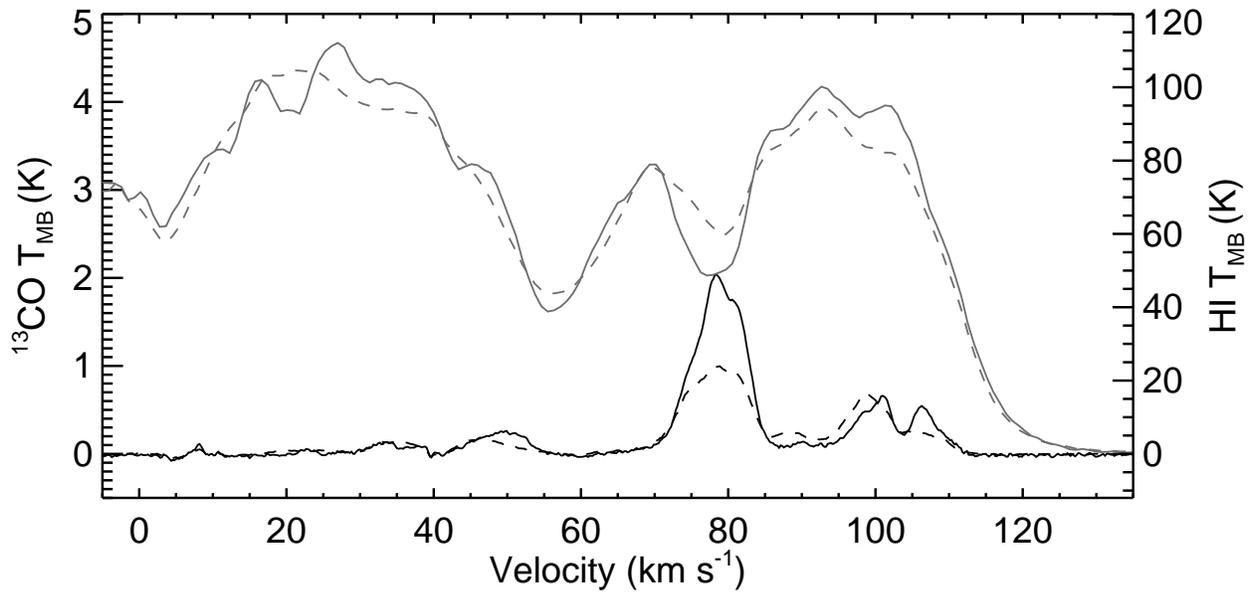}

\caption{Average on--(solid curves) and off--cloud (dashed curves) spectra
 of \hi\ (gray curves) and \cor (black curves) for \irdci.}

\label{fig:hisa}
\end{figure}

Figure~\ref{fig:hisa} shows the average on-- (solid curves) and
off--cloud (dashed curves) spectra of \cor\ (black curves) and
\hi\ (gray curves) for Figure~\ref{fig:hisa}.  For \irdci,
\hi\ absorption is seen at the velocity of all \cor\ components with
velocities less than the cloud velocity of 78.6 \kms (this is clearer
when analyzing each clump individually).  Therefore, we assign all
\cor\ emission not associated with \irdci\ that has velocities less
than 76.6\,\kms\ to the near component.  We assign all \cor\ features
with velocities greater than 78.6\,\kms\ to the far component.  Also
evident in Figure~\ref{fig:hisa} is the strong \hi\ absorption at the cloud
velocity, which shows clearly that the cloud is at the near distance.
The \hi\ emission is less strongly peaked than the \cor\ emission.

The results of our column density modeling are summarized in
Table~\ref{tab:column}.  This table gives the column density of atomic
hydrogen, $N({\rm H})$, molecular hydrogen $N(\rm H_2)$, and the total
column density, $N_{\rm Total} = N({\rm H}) + 2 N(\rm H_2)$ in units
of $\expo{22}$\persqcm\ for \irdci\ and \irdcii.  The contributions to
the column density are given for the foreground and background
components for the on-- and off--cloud directions.  The background
component includes the contributions from the IRDCs themselves.

\section{Spectral Analysis \label{sec:spectra}}
Using our cleaned events files, we extract the X-ray spectra for 
both the on-- and off--cloud regions in preparation for spectral fitting.
Our on-- and off--cloud regions are defined as before and correspond
to high infrared extinction and high \cor\ column density.  We extract
spectra from both MOS detectors and create auxiliary response files
(ARFs), redistribution matrix files (RMFs), and background spectra.
We increase the signal to noise by binning these spectra so there are
at least 20 counts per spectral channel.  We use XSPEC\footnote[4]{See
  http://xspec.gsfc.nasa.gov/docs/xanadu/xspec} V11.3.2
\citep{arnaud96} for the spectral analysis.

For the spectral fits, we assume the emission can be described as:
\begin{equation}
I_X = SP + I_{LHB} + I_F + I_B + (I_R + AGN)\,,
\label{eq:compts}
\end{equation}
where $I_X$ is the measured X-ray flux, $SP$ is the contribution from
soft proton contamination, $I_{LHB}$ is the emission from the LHB,
$I_F$ is the Galactic X-ray emission foreground to the cloud, $I_B$ is
the Galactic X-ray emission background to the cloud, $I_R$ is the
emission from the Galactic ridge, and $AGN$ is the X-ray background
from unresolved active galactic nuclei.
There are three column density parameters in our models, one
associated with $I_F$, one with $I_B$, and one with both $I_R$ and
$AGN.$  The fitted column density associated with $I_B$ is poorly constrained
(see below), which is why we do not treat the column density
associated with $I_R$ and $AGN$ as the sum of that affecting $I_F$ and
$I_B$.  Below we describe in more detail the various model components
we use to model the X-ray background and to separate foreground from
background emission.

There are two main sources of contamination in our spectra.  First,
there is residual soft proton contamination not removed through time
filtering ($SP$ in Equation~\ref{eq:compts}).  This component should
have a power law distribution of energies, so we model its emission
using the XSPEC model {\tt pow/b}\footnote{The notation for this model
  is that of XSPEC~V11; in more recent versions the ``/b'' has been
  dropped.  See
  http://heasarc.gsfc.nasa.gov/docs/xanadu/xspec/backgroundmodel.html.}.
This XSPEC component models a power-law noise distribution not folded
through the instrument response.  Secondly, there are two fluorescent
instrumental lines of Al K$\alpha$ and Si K$\alpha$ near 1.6 \kev.  We
assume a Gaussian shape for these lines and fit them explicitly.

Our models include three components for the X-ray background, as in
\citet{snowden00}.  Along each line of sight, first there is $\sim
0.1$ \kev\ emission from the LHB \citep{smith01}, $I_{LHB}$.  We fit
the the LHB emission with the XSPEC model {\tt apec} \citep{smith01b},
which represents collisionally-ionized diffuse plasma.  Next, we
assume that there are two components of the X-Ray background emission
in the Galactic plane: one soft ($\lesssim$ 1 \kev) and one hard (few
\kev), $I_R$.  For these components we use the XSPEC model {\tt wabs}
$\times$ {\tt apec}, which represents collisionally-ionized diffuse
plasma attenuated by material along the line of sight.  The soft
component itself can be divided into two sources: emission in front of
the IRDC, $I_F$, and emission from beyond the IRDC, $I_B$.  Finally,
there is extragalactic emission from unresolved AGN in the field,
$AGN$.  We model the contribution from AGN using a power law model
with absorption, {\tt wabs} $\times$ {\tt pow}, where the power law
index is set to $\alpha = 1.46$ \citep{chen97}.  This is expected to
be the dominant component above 1 \kev\ \citep{lumb02}.

\begin{figure}
\includegraphics[width=6.5in]{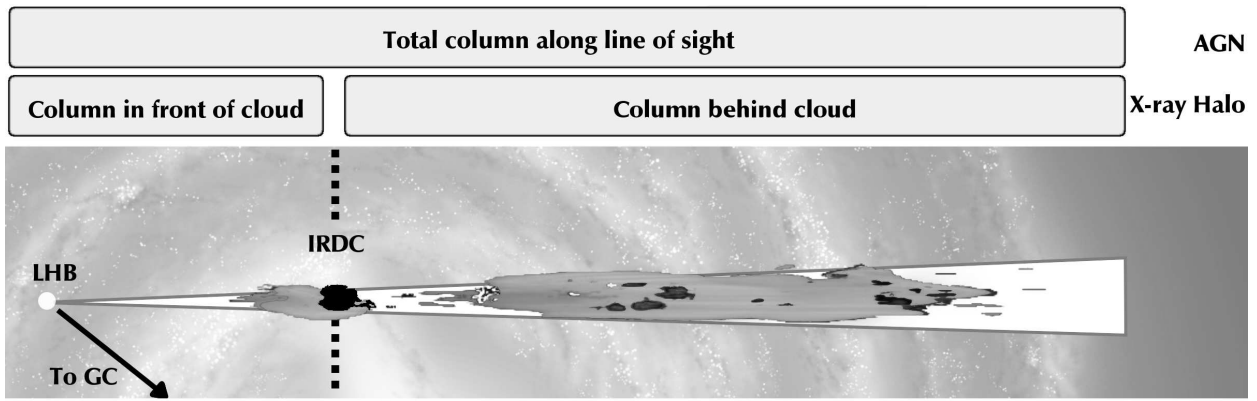}

\caption{Components of the X-ray background attenuated by the Galactic
column density for \irdci.  The column density Molecular gas from the GRS is shown black
while atomic gas from the VGPS is shown in gray.  The Local Hot Bubble
(labeled LHB), the direction of our observation (white wedge) and the
direction towards the Galactic center are also shown.  The size of the
wedge and the LHB are slightly exaggerated for clarity.  IRDC
\irdci\ is associated with a large dense molecular cloud 5 kpc from
the Sun.)}

\label{fig:cartoon}
\end{figure}

We show the model components and the column densities affecting them
graphically in Figure~\ref{fig:cartoon} for \irdci.  In this figure,
molecular gas is displayed in black and atomic gas is displayed in
gray.  The white wedge is an exaggerated representation of the \xmm\
field of view.  Also shown on the figure are the location of IRDC
\irdci, the LHB, and the direction to the Galactic
center (GC).  The LHB size in the figure is exaggerated for clarity.
Figure~\ref{fig:cartoon} shows that the AGN X-ray emission is affected
by the total Galactic column density.

During the spectral modeling, we fit the spectra for each cloud
simultaneously to improve the fit quality.  We fix the abundance to
Solar and use the Solar relative metal abundance of \citet{anders89}.
We have five spectra for each cloud: on-- and off--cloud spectra for
both MOS detectors, as well as a ROSAT All Sky Survey
\citep[RASS;][]{snowden97} spectrum derived from the HEASARC X-ray
Background
Tool.\footnote[5]{http://rosat.gsfc.nasa.gov/cgi-bin/Tools/xraybg/xraybg.pl}
The RASS spectrum helps constrain the cosmic background at lower
energies where the \xmm\ detectors are less sensitive.  We extract the
RASS spectra from regions centered on our clouds of radius
$0.5\arcdeg$ and normalize the spectra to 1 arcminute$^2$.  We use the
same RASS spectrum for the on-- and off--cloud regions, but scale the
emission by the relative area of these regions.  We link the
normalizations and temperatures of each spectral component for all
five spectra to constrain the fits.  We also link the temperature of
the soft foreground component to that of the soft background
component.

\begin{figure}
\includegraphics[width=6.5in]{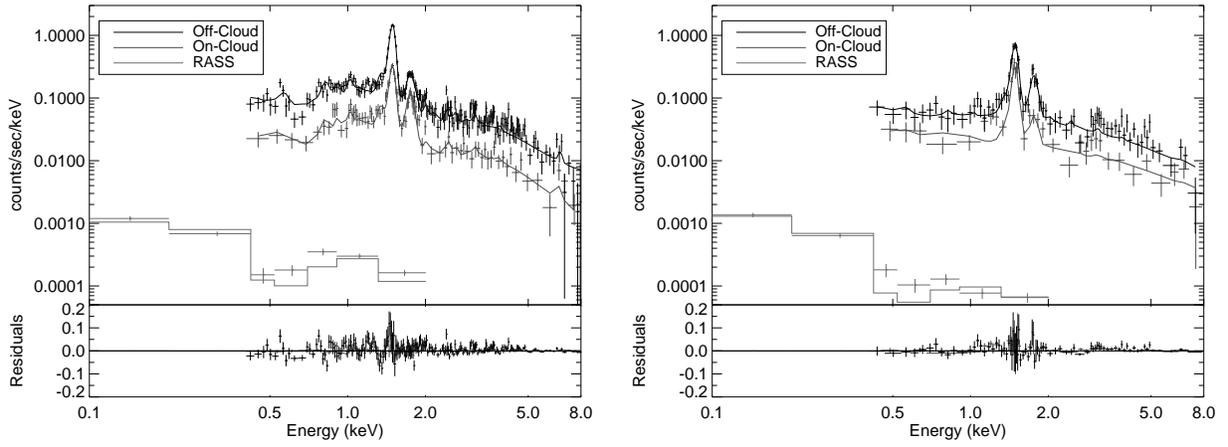}

\caption{Spectral data and models for \irdci\ (left panel, column
  density free fit) and \irdcii\ (right panel) from the MOS1 detector.
  Shown are the on--cloud (black), off--cloud (dark gray), and RASS
  (light gray) spectra.  The bottom section of each panel shows the
  fit residuals.}

\label{fig:spectra}
\end{figure}

The results of our spectral fitting procedure are shown in
Figure~\ref{fig:spectra} for the MOS1 detector.  In
Figure~\ref{fig:spectra}, the top curve is the on--cloud spectrum, and
the lower curve is the off--cloud spectrum.  The bottom curve is the
RASS spectrum.  The two prominent emission lines are the fluorescent
instrumental lines of Al K$\alpha$ and Si K$\alpha$.  The left panel
of Figure~\ref{fig:spectra} shows the spectra and model for \irdci,
and the right panel shows the same for \irdcii.  It is evident from
this figure that our model adequately fits the data.  Also apparent
from Figure~\ref{fig:spectra} are the decreased counts for \irdcii\ due
to the soft proton contamination.  This results in greater uncertainty
in the spectral fits.

For \irdci, we run our fits in two trials, once with the column
densities fixed to their calculated values (see \S\ref{sec:column})
and once with the column densities allowed to vary (hereafter the
``fixed'' and ``free'' trials).  We allow the column densities to vary
up to 100\% from our calculated values to account for uncertainties in
converting line parameters to column densities.  Due to insufficient
counts, we are unable to allow the column densities to vary for
\irdcii, and thus we run the spectral fits only with fixed values.

\subsection{Spectral Fit Results}
\begin{figure}
\includegraphics[width=6.5in]{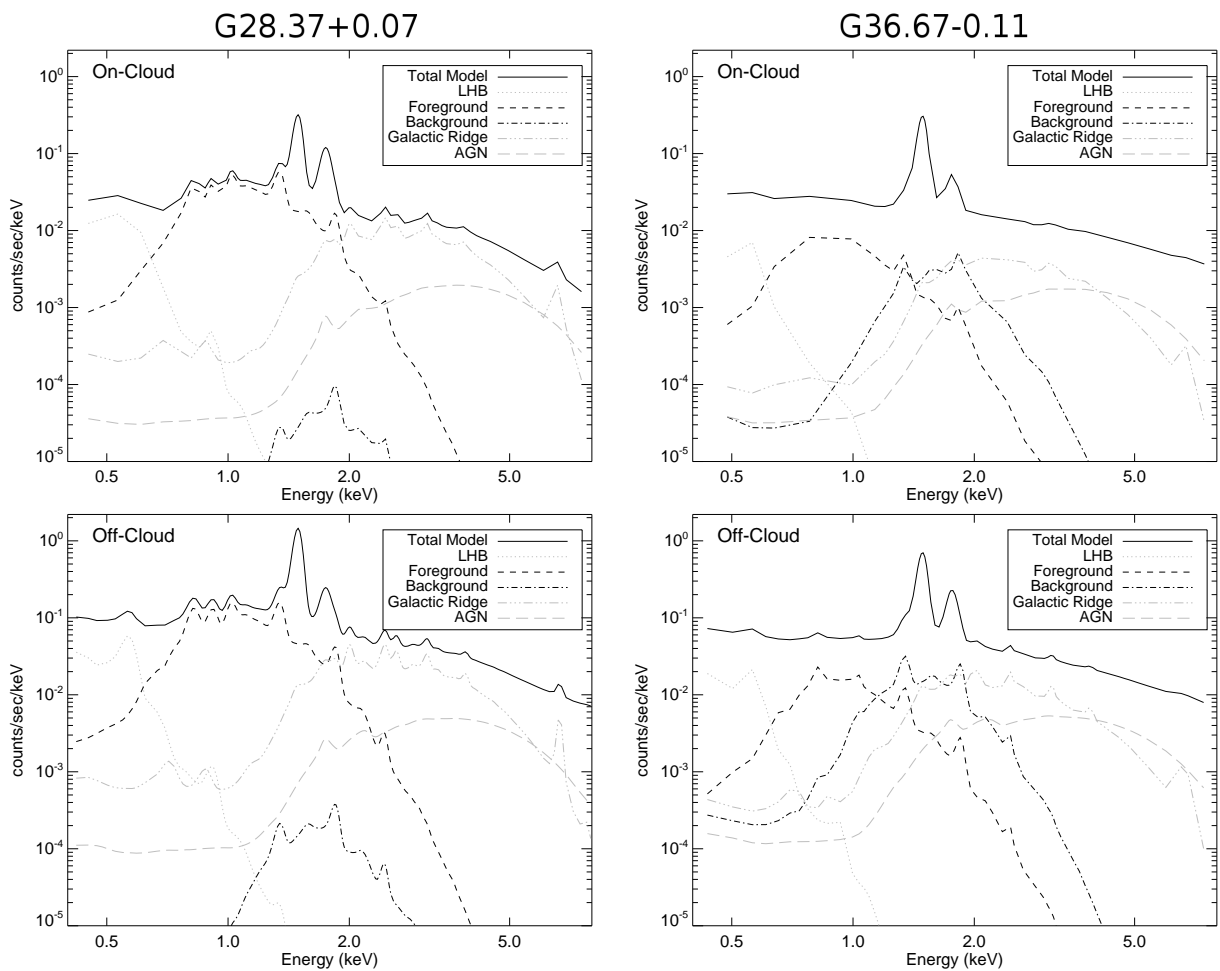}

\caption{Model spectral components from the XSPEC fit to the on--cloud
  (top panels) and off--cloud (bottom panels) spectra for \irdci\
  (left panels) and \irdcii\ (right panels) from the MOS1 detector.
  For \irdci, the models are from the trial with column density free.}

\label{fig:model}
\end{figure}

The results of our spectral fits are shown in Figure~\ref{fig:model}
and listed in Table~\ref{tab:spectra}.  Figure~\ref{fig:model} shows
the best fit model components for \irdci\ (left panels) and \irdcii\
(right panels).  The top panels show the models for the on--cloud
directions, and the bottom panels show the same for the off--cloud
directions.  The models for \irdci\ are from the free trial.
Table~\ref{tab:spectra} lists the fitted temperature and emission
measure (EM = $\int n_e^2 dl$) for our four spectral model components,
the reduced $\chi^2$ value, and the number of degrees of freedom in
the fit.  Errors in the table represent 90\% confidence levels.  Below
we discuss the model components individually.

\subsubsection{Column Densities}
When allowed to vary, the column densities for \irdci\ fit to similar,
but generally lower, values compared to our calculated values (see \S
\ref{sec:column}).  For the on--cloud spectra, the column density fits
to $1.2^{+0.03}_{-0.02} \times \expo{22}$ \persqcm\ for the foreground
component; we estimated $1.3\times \expo{22}$ \persqcm. The column
density background to the cloud is unconstrained when allowed to vary
as its normalization fits to zero (see \S \ref{sec:bgcomp} below).
The total column density affecting the Galactic ridge and AGN emission
fits to $3.9^{+0.6}_{-0.5} \times \expo{22}$ \persqcm\ for the
on--cloud spectra; we estimated $5.2\times \expo{22}$ \persqcm.
Thus, since the total column density affecting the Galactic Ridge and
AGN component is the sum of that foreground and background to the
cloud, we find $2.6\times \expo{22}$ for the background component.
For the off--cloud spectra, the column density fits to $1.1
^{+0.04}_{-0.03}\times \expo{22}$ \persqcm\ foreground to the cloud
and to $2.9 ^{+0.3}_{-0.2}\times \expo{22}$ \persqcm\ for the total
column density; we estimated $0.9\times \expo{22}$ \persqcm\ and
$4.2\times \expo{22}$ \persqcm, respectively.

Perhaps the most important column density parameter -- the difference
in column density between the on- and off--cloud regions -- is the
same in the fixed and free trials for \irdci: $1.0 \times \expo{22}$
\persqcm, although with a formal error derived from the fit of about
50\%.  The other discrepancies are discouraging, although the
differences have little affect on the other model parameters.  The
largest effect can be seen in the emission measure of the foreground
soft component (see below).  Despite our care taken in estimating the
column densities, errors in the conversion from \cor\ line parameters
to hydrogen column density are large and currently unavoidable.
Longer or more sensitive observations are required to more accurately
disentangle the competing effects of absorption from the Galactic
column density and emission from the X-ray background.

\subsubsection{LHB Component}
The fixed and free column density trials for \irdci\ both result in a
LHB component temperature $kT = 0.11$ \kev\ (log$T$/K = 6.11).  For
\irdcii, the LHB component fits to a temperature of 0.089 \kev
(log$T$/K = 6.02.  These temperatures for the LHB component
are in good agreement with that found by many previous studies
\citep[see][]{snowden98, snowden00, kuntz00, henley07}.  The emission 
measure for the LHB component is less than what has been found by 
previous authors.  We calculate an emission measure of 
$5.5\times 10^{-4}$ cm$^{-6}$ pc for the fixed trial and
$5.4\times 10^{-4}$ cm$^{-6}$ pc for the free trial for \irdci.  For \irdcii, it
is slightly higher: $6.3\times 10^{-4}$ cm$^{-6}$ pc.  These values are three times
less than that found by \citet{henley07}, $\sim 10$ times less than
that found by \citet{smith07}.  This emission measure is in large part 
determined from the RASS 1/4 keV band data, and as the directions lie 
in the Galactic plane the surface brightness, and therefore the emission
measure of the RASS data are relatively low.  

\subsubsection{Soft Foreground Component}
The soft component fits to a temperature of 0.29 \kev\ (log$T$/K =
6.53) in the fixed trial and to 0.30 \kev\ (log$T$/K = 6.54) in the
free trial for \irdci.  For \irdcii, the soft component is not very
well constrained.  In fact, the only constraint we can place on this
component is that it is less than 0.35 \kev (log$T$/K = 6.61).
\citet{kuntz08} find a similar temperature for this component: 0.24
\kev.

\subsubsection{Soft Background Component \label{sec:bgcomp}}
For the fixed trial of \irdci, the soft background component has
little effect on the model.  In fact, its removal has no impact on the
quality of the fits, nor any impact on the value of the other
parameters.  The large column density in front of the background
component causes significant absorption.  The negligible effect this
component has on our total model can be seen in
Figure~\ref{fig:model}.  For the free trial, the normalization for
this component fits to zero.  For \irdcii, the soft foreground
component and soft background component have similar intensities.  The
difference between the fits seen in Figure~\ref{fig:model} for the two
clouds is due partly to the decreased column density to \irdcii.
Perhaps a larger factor, however, could be the greater uncertainty in
the spectral fits for \irdcii\ due to soft proton contamination.

\subsubsection{Galactic Ridge Component}
The temperature for the warm Galactic ridge component fits to 1.61
\kev\ (log$T$/K = 7.27) in the free trial and to 2.34 \kev\ (log$T$/K
= 7.43) in the fixed trial for \irdci.  This is a significantly higher
than what has been found by previous authors.  \citet{kuntz08} find a
temperature of 0.71 \kev\ for this component, with rather large error
bars.  Because of the decreased counts in the observation of \irdcii,
we were forced to fix the Galactic ridge component to the temperature
found for \irdci.  When allowed to vary, this component would fit to
an extremely high temperature ($> 10\, \kev$), which we deem
unphysical, especially considering the results for \irdci.

Figure~\ref{fig:model} shows that for both clouds the Galactic ridge
emission is the strongest component of the model between 2 and 5 \kev.
To determine if this component is required by the spectral fits, we
remove the hard component from the models for \irdci.  Its exclusion
results in significantly worse fits: the reduced $\chi^2$ value
increases from 1.11 to 1.41.  An F-test reveals that the inclusion of
the Galactic ridge component is significant at the $15\sigma$ level.
We conclude that this is a necessary component of our spectral fits
although its temperature is not very well constrained.

\subsection{O VII and O VIII Emission}
Oxygen is the most abundant line-emitting element at the temperatures
of the LHB.  In a million degree plasma, oxygen is primarily in the
O$^{+6}$ state (O\,{\sc vii}), and the strongest lines detectable by
\xmm\ are near 0.5 \kev.  The emission from other charge states (most
important for our purposes is O\,{\sc viii}) can be used as a
diagnostic to constrain the temperature of the LHB and verify the
temperature found previously through spectral fitting.  There are
three lines of O\,{\sc vii} that are all near 0.57 \kev, and there is
a Ly$\alpha$ transition of O\,{\sc viii} at 0.65 \kev.

We measure the intensities of the O\,{\sc vii} and O\,{\sc viii}
emission from the LHB by replacing the {\tt apec} model
component with a {\tt vapec} model.  The {\tt vapec} model is
identical to the {\tt apec} model except that the atomic abundances
may be individually modified.  We set the oxygen abundance to zero and
add two Gaussians representing the O\,{\sc vii} and O\,{\sc viii}
emission lines at 0.57 \kev\ and 0.65 \kev.  For \irdci, when we fix
all the model parameters to their previously fit values (except for
the {\tt vapec} temperature and normalization), we find that the
O\,{\sc vii} line has an intensity of 3.4$^{+0.4}_{-1.0}$ photons
cm$^{-2}$ s$^{-1}$ sr$^{-1}$ (hereafter line units, LU).  The {\tt
  vapec} temperature fits to a value of 0.098 \kev.  When we fix the
{\tt vapec} temperature to the LHB temperature found previously, 0.11
\kev, we find an O\,{\sc vii} intensity of 3.3$^{+0.7}_{-0.8}$ LU.  We
do not detect the O\,{\sc viii} line in either of these trials.  The
same procedure applied to \irdcii\ yields similar results.  With the
{\tt vapec} temperature allowed to vary, the O\,{\sc vii} intensity is
$1.9^{+1.3}_{-1.6}$ LU.  The {\tt vapec} temperature fits to 0.085
\kev.  When fixed to the 0.089 \kev\ value found previously, the
O\,{\sc vii} intensity is $2.1^{+1.3}_{-1.3}$ LU.  We do not
detect the O\,{\sc viii} line.

Our values for the O\,{\sc vii} and O\,{\sc viii} lines are in rough
agreement with what has been found by other authors.  In a shadowing
experiment to a nearby ($\sim 200$ pc) filament,
\citet{henley07} found $3.4^{+0.6}_{-0.4}$ LU for the O\,{\sc vii}
line and did not detect the O\,{\sc viii} line.  \citet{kuntz08} found
$1.75\pm0.7$ LU for the O\,{\sc vii} emission at ({\it l,b}\/) =
($111\fdeg14, 1\fdeg11$).  In a shadowing experiment of MBM~12,
a high latitude cloud $\sim 50$ to $\sim 300$ pc distant,
\citet{smith05} found $1.79\pm 0.55$ LU for O\,{\sc vii} and
$2.34\pm0.36$ LU for O\,{\sc viii}.  They note that this unusually
strong O\,{\sc viii} emission can be explained by solar wind charge
exchange (SWCX) contamination in their spectra.  The SWCX emission is
caused by electron transitions between neutral atoms and highly
ionized species in the solar wind.  It has a potentially large effect
in X-ray data, especially following a coronal mass ejection event.
\citet{snowden04} showed the large effect on the oxygen lines that
SWCX emission can have.  In a followup study of MBM~12 with {\it Suzaku}, 
\citet{smith07} found $2.34\pm0.33$ LU for O\,{\sc vii} and
$0.77\pm0.16$ LU for O\,{\sc viii} line emission.  Using the X-ray
Quantum Calorimeter aboard a sounding rocket,
\citet{mccammon02} found $4.8\pm0.8$ LU for the O\,{\sc vii} emission
and $1.6\pm0.4$ LU for the O\,{\sc viii} emission.

While consistent with previous values, we note that there is likely
some residual contamination from SWCX in our measured oxygen line
intensities.  Using the models in \citet{smith05}, we estimate that for
the EM and $T$ found here for the LHB, we should expect O\,{\sc
  vii} and O\,{\sc viii} intensities of $\sim2$ and $<1$ LU, respectively.  This
indicates that the line intensities measured here cannot be solely due
to a plasma in thermal equilibrium.

A consensus appears to be emerging that most of the O\,{\sc vii}
emission arises from the LHB (with a contribution from SWCX).  Strong
O\,{\sc viii} emission, however, cannot arise from the LHB as this
would imply a temperature inconsistent with other determinations.
This emission must arise from SWCX \citep{snowden04} or from a cool
component distributed throughout the Galaxy \citep{kuntz08}, likely
with contributions from both sources.  Our results are consistent
with this interpretation.
That we do not detect the O\,{\sc viii} line is not
surprising given the large absorbing column density and low expected line
strength.

\section{Image Analysis}
We produce exposure corrected, background subtracted X-ray images to
verify the results from the spectral analysis, and to search for a
morphological match between the molecular emission and an X-ray
decrement.  Using the XMM-ESAS software, we create these images in
three energy bands: 0.35--1.25 \kev, 1.25--2.00 \kev, and 2--10 \kev.
We model two sources of background in our images: the particle
background and residual soft proton contamination.  These background
components must be removed as cleanly and completely as possible to
ensure accurate results.

Using the filter-wheel-closed data, the XMM-ESAS software models the
particle background \citep[see][]{kuntz08}.  These data are dominated
by the instrumental background because the chips were not exposed to
the sky.  We do not include blank sky data in the modeling of either
the particle background nor the soft proton contamination.  The blank
sky files may suffer from residual soft proton contamination and solar
wind charge exchange.  Additionally, the blank sky files contain the
cosmic background which we are attempting to observe.  Using these
data would not only add error to our models, but would also remove the
signal of interest.

The XMM-ESAS software also allows for the creation of a model image of any
residual soft proton contamination.  \citet{kuntz08} characterize the
reasonable ranges of the variations in both the spectrum and in the
spatial distribution over the detectors.  XMM-ESAS incorporates this
characterization in the model soft proton image.  The
energy distribution of the soft proton contamination is determined from
the power law index and normalization found in the spectral fits.  For
this input, we use the values found in the spectral fitting.

\begin{figure}
\includegraphics[scale=0.75, clip]{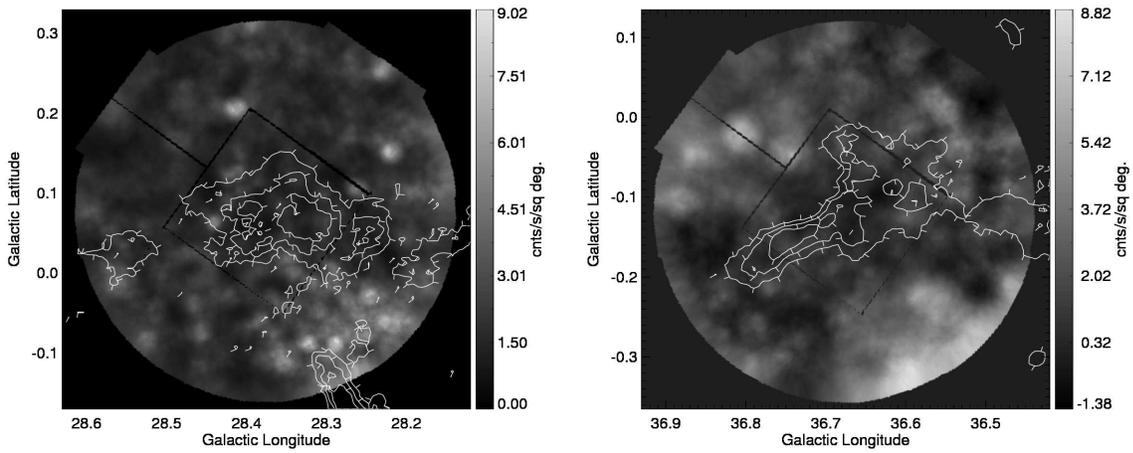}

\caption{Smoothed \xmm\ images in the energy range 0.35 \kev\ to 1.25
  \kev\ of \irdci\ (left panel) and \irdcii (right panel).  The
  contours are the same GRS contours shown in Figure~\ref{fig:clouds}.
  There is no obvious X-ray decrement at the location of the molecular
  clouds.}

\label{fig:smooth}
\end{figure}

We subtract the model background and soft proton images, combine the
data from the two MOS detectors and smooth the resultant image.  The
result of these operations is an exposure corrected, background
subtracted image.  The strongest X-ray shadow should be present in the
0.35--1.25 \kev\ image.  We show in Figure~\ref{fig:smooth} the exposure
corrected, background subtracted image for the 0.35-1.25 \kev\ band of
\irdci\ (left panel) \irdcii\ (right panel).
In the images of both clouds there is some
indication of a shadow, but it not the clearly defined shadow one
would expect, nor does it have the same depth one would expect.
Furthermore, the same shadow that appears in the 0.35--1.25 band also
appears in the 2.0--8.0 \kev\ band.  As there should not be strong
absorption in the 2.0--8.0 \kev\ band, this indicates that the
``shadow'' is a product of the detector, and not of the cloud itself.
We conclude that the imaging analysis supports the spectral analysis
and finds no X-ray shadow of the cooler component to the X-ray
background.

\section{Summary}
We have conducted X-ray shadowing experiments on two infrared dark
clouds near $l = 30\arcdeg$ at distances of 2 and 5 kpc from the Sun
to determine the spatial distribution of the X-ray background.  We
used \hi\ and \cor\ data to calculate the total atomic and molecular
column densities for both the on-- and off--cloud directions.  Both
clouds have very high proton column densities of $\gtrsim 4 \times
\expo{22}\, {\rm cm^{-2}}$.  This high column density should absorb
nearly all the soft background X-ray flux.  We find that the cool,
diffuse X-ray background must originate foreground to the clouds,
within a few kpc of the Sun.  Our results are further evidence that
X-ray emitting plasma is distributed throughout the disk of our Galaxy
\citep[see][]{park97, kuntz08}.




We found that the X-ray background is best fit with a three component
model with contributions from the Local Hot Bubble (LHB), a soft
component, and a hot component from the Galactic ridge.  The LHB
emission is best fit with a value of $\sim 0.1$ \kev\ (log $T$/K =
6.06) for both clouds, in agreement with what has been found in
previous studies.  The soft component is best fit with a temperature
of $\sim 0.3$ \kev\ (log $T$/K = 6.54).  This is the dominant source
of emission between 0.7 and 1.0\,\kev.  The Galactic ridge component is
best fit with a temperature of 2\,\kev.  This is the dominant source of
emission in our spectral models between 2 and 5 \kev.  Our spectral
fits show that most significant emission below $\sim0.7$\,\kev\ can be
attributed to the LHB (and/or to SWCX).  Image analysis including
accurate modeling of any background components failed to reveal an
X-ray shadow, supporting the results of the spectral modeling.

We detect the O\,{\sc vii} line for both clouds at levels of $\sim\,3$
photons cm$^{-2}$ s$^{-1}$ sr$^{-1}$.  This intensity is $\sim50\%$
higher than what one would expect from a thermal plasma with the emission
measure and temperature of the LHB, indicating some level of
contamination from solar wind charge exchange.  These values are
roughly in agreement with what has been found in previous studies.  We
do not detect the O\,{\sc viii} line for either cloud.

\begin{acknowledgements}
  This research was based on observations obtained with {\it
    XMM-Newton}, an ESA science mission with instruments and
  contributions directly funded by ESA Member States and NASA, and was
  supported by NASA {\it XMM-Newton} Guest Observer grants including
  NNX06AG73G and NNG05GP68G.  We make use of molecular line data from
  the Boston University-FCRAO Galactic Ring Survey (GRS). The GRS is a
  joint project of Boston University and the Five College Radio
  Astronomy Observatory (FCRAO), funded by the National Science
  Foundation under grants AST-9800334, AST-0098562, AST-0100793,
  AST-0228993, \& AST-0507657.  We also make use of data from the VLA
  Galactic plane survey (VGPS).  The National Radio Astronomy
  Observatory is a facility of the National Science Foundation
  operated under cooperative agreement by Associated Universities,
  Inc.  We would like to thank the referee for their careful reading
  of this paper and their helpful comments which greatly improved its
  clarity.
\end{acknowledgements}

\clearpage

\clearpage

\clearpage


\clearpage

\clearpage

\begin{deluxetable}{lccccc}
\tabletypesize{\scriptsize}
\tablecaption{Infrared Dark Cloud Properties}
\tablewidth{0pt}
\tablehead{
\colhead{Name} & 
\colhead{l} & 
\colhead{b} &
\colhead{D} &
\colhead{Size} &
\colhead{N(H$_2$)} \\

\colhead{} & 
\colhead{deg.} & 
\colhead{deg.} & 
\colhead{kpc} &
\colhead{sq. arcmin} &
\colhead{$10^{22}$ cm$^{-2}$}
}
\startdata
\irdci\ & 28.37 & +0.07 & 5.0 & 42 & 21 \\
\irdcii\ & 36.67 & $-$0.11 & 2.0 & 19 & 19 \\
\enddata
\label{tab:prop}
\end{deluxetable}

\begin{deluxetable}{lccccccccc}
 \tabletypesize{\scriptsize}
 \tablecaption{Cloud Column Densities$^\dag$}
 \tablenotetext{\dag}{All column densities are in units of \expo{22}\persqcm. 
   ``FG'' and ``BG'' represent foreground and background to the cloud.}
 \tablewidth{0pt}
 \tablehead{
   \colhead{Name} &
   \multicolumn{2}{c}{$N({\rm H})$} &
   \colhead{} &
   \multicolumn{2}{c}{$N({\rm H}_2)$} &
   \colhead{} &
   \multicolumn{2}{c}{$N({\rm Total})$} \\  \cline{2-3} \cline{5-6} \cline{8-9}

   \colhead{} &
   \colhead{FG} &
   \colhead{BG} &
   \colhead{} &
   \colhead{FG} &
   \colhead{BG} &
   \colhead{} &
   \colhead{FG} &
   \colhead{BG}
 }
 \startdata
\irdci\ \\
\ \ \ \ On--cloud & 0.5 & 1.0 & & 0.4 & 2.1 & & 1.3 & 5.2 \\
\ \ \ \ Off--cloud & 0.5 & 1.0 & & 0.2 & 1.6 & & 0.9 & 4.2 \\
\irdcii\ \\
\ \ \ \ On--cloud & 0.5 & 1.0 & & 0.1 & 1.4 & & 0.7 & 3.8 \\ 
\ \ \ \ Off--cloud & 0.5 & 1.0 & & 0.1 & 0.9 & & 0.7 & 2.8 \\

 \enddata
 
 \label{tab:column}
\end{deluxetable}
\clearpage

\begin{deluxetable}{lcccccccc}
 \tabletypesize{\scriptsize}
 \tablecaption{Spectral Model Parameters}
 \tablewidth{0pt}
 \tablehead{
 \colhead{} & \multicolumn{5}{c}{\irdci} & & \multicolumn{2}{c}{\irdcii} \\ \cline{2-6} \cline{8-9} 
 \colhead{} & \multicolumn{2}{c}{Column Fixed} & & \multicolumn{2}{c}{Column Free} & & 
    \multicolumn{2}{c}{Column Fixed} \\ \cline{2-3} \cline{5-6} \cline{8-9}
 \colhead{Parameter} & \colhead{Value} & \colhead{Error} & & \colhead{Value} & \colhead{Error} & &
    \colhead{Value} & \colhead{Error}
 }
\startdata

Column Density $\times 10^{22}$ cm$^{-2}$ & & & & & & \\
\ \ \ \ On--cloud, Foreground & 1.3 & \nodata & & 1.2 & $^{+0.003}_{-0.02}$ & & 0.7 & \nodata \\
\ \ \ \ On--cloud, Background & 5.2 & \nodata & & 3.9 & $^{0.6}_{-0.5}$ & & 3.8 & \nodata \\
\ \ \ \ Off--cloud, Foreground & 0.9 & \nodata & & 1.1 & $^{0.04}_{-0.03}$ & & 0.7 & \nodata \\
\ \ \ \ Off--cloud, Background & 4.2 & \nodata & & 2.9 & $^{0.3}_{-0.2}$ & & 2.8 & \nodata \\
LHB & & & & & & \\
\vspace{0.04in} \ \ \ \ kT(keV) & 0.11 & $^{+3.7{\rm E}-3}_{-9.3{\rm E}-3}$ & & 0.11 & $^{+8.3{\rm E}-3}_{-9.7{\rm E}-4}$ & & 
	0.089 & $^{+4.7{\rm E}-3}_{-9.2{\rm E}-3}$ \\ \vspace{0.04in}
\ \ \ \ EM(cm$^{-6}$ pc) & 5.5E$-$4 & $^{+7.6{\rm E}-5}_{-7.0{\rm E}-5}$ & & 5.4E$-$4 & $^{+6.0{\rm E}-5}_{-7.1{\rm E}-5}$ & & 
	6.3E$-$4 & $^{+3.6{\rm E}-5}_{-6.8{\rm E}-5}$ \\
Cool Foreground & & & & & & \\
\vspace{0.04in} \ \ \ \ kT(keV) & 0.29 & $^{+1.5{\rm E}-2}_{-7.6{\rm E}-3}$  & & 0.30 & $^{+2.8{\rm E}-2}_{-1.5{\rm}-2}$ & & 
	$<0.35$ & \nodata \\
\ \ \ \ EM(cm$^{-6}$ pc) & 4.6E$-$2 & $^{+6.9{\rm E}-3}_{-9.8{\rm E}-3}$ & & 3.4E$-$2 & $^{+2.5{\rm E}-3}_{-1.0{\rm E}-2}$ & & 
	$2.3{\rm E}-3$ & $^{+3.7{\rm E}-4}_{-1.5{\rm E}-3}$ \\
Cool Background & & & & & & \\
\vspace{0.04in} \ \ \ \ kT(keV) & 0.29 & \nodata & & 0.30 & \nodata & & 
	$<0.35$ & \nodata \\
\ \ \ \ EM(cm$^{-6}$ pc) & 1.9E$-$3 & $^{+1.6{\rm E}-1}_{-1.9{\rm E}-3}$ & & 0.00 & $^{+3.9{\rm E}-5}_{-0.0}$ & & 
	5.8E$-$2 & $^{+1.9{\rm E}-2}_{-8.7{\rm E}-3}$ \\
Galactic Ridge & & & & & & \\
\vspace{0.04in} \ \ \ \ kT(keV) & 1.61 & $^{+0.25}_{-0.14}$ & & 2.34 & $^{+0.35}_{-0.28}$ & & 
	1.61 & \nodata \\
\ \ \ \ EM(cm$^{-6}$ pc) & 2.2E$-$2 & $^{+1.5{\rm E}-3}_{-4.7{\rm E}-3}$ & & 1.2E$-$2 & $^{+2.1{\rm E}-3}_{-1.1{\rm E}-3}$ & & 
	8.2E$-$3 & $^{+1.6{\rm E}-3}_{-2.6{\rm E}-3}$ \\
$\chi^2$ & 1373.5 & \nodata & & 1348.1 & \nodata & & 
	352.4 & \nodata \\
dof & 1236 & \nodata & & 1231 & \nodata & & 
	300 & \nodata \\

\enddata

\label{tab:spectra}
\end{deluxetable}

\end{document}